\documentclass[twocolumn,showpacs,pre]{revtex4}

\usepackage{graphicx}
\usepackage{amssymb}

\begin{document}

\title{Probing tails of energy distributions using importance-sampling
in the disorder\\ with a guiding function}

\author{Mathias K\"orner} 
\author{Helmut G.~Katzgraber} 
\affiliation{Theoretische Physik, ETH Z\"urich, CH-8093 Z\"urich, Switzerland}

\author{Alexander K.~Hartmann}
\affiliation{Institut f\"ur Theoretische Physik, Universit\"at G\"ottingen, 
Friedrich-Hund-Platz 1, 37077 G\"{o}ttingen, Germany}

\date{\today}

\begin{abstract}

We propose a simple and general procedure based on a recently introduced 
approach that uses an importance-sampling Monte Carlo algorithm in the 
disorder to probe to high precision the tails of ground-state energy 
distributions of disordered systems. Our approach requires an estimate of
the ground-state energy distribution as a guiding function which can be
obtained from simple-sampling simulations.
In order to illustrate the algorithm, we compute the ground-state energy
distribution of the Sherrington-Kirkpatrick mean-field Ising spin glass to
eighteen orders of magnitude. We find that if the ground-state energy 
distribution in the thermodynamic limit is described by a modified Gumbel 
distribution as previously predicted, then the value of the slope parameter 
$m$ is clearly larger than $6$ and of the order $11$.
\end{abstract}

\maketitle

\section{Introduction}
\label{sec:introduction}

The ground-state energy distributions of disordered systems and in particular
of spin glasses \cite{binder:86,mezard:87,young:98,diep:05} have recently 
received considerable attention from different
groups. Bouchaud \textit{et~al.}~\cite{bouchaud:03} have studied the
ground-state energy distribution of short-range spin glasses in two and three
dimensions and find them to become Gaussian in the thermodynamic 
limit \cite{aizenman:90}, while
other groups \cite{palassini:03a,andreanov:04,boettcher:05,boettcher:05a}
find that the ground-state energy distribution
of the mean-field Sherrington-Kirkpatrick (SK) model \cite{sherrington:75} 
remains skewed in the thermodynamic limit and apparently can be
described by a modified Gumbel distribution \cite{gumbel:60}.
In a recent paper, we have been able to show that this transition from a 
Gaussian to a non-Gaussian limiting distribution coincides with the 
transition from long-range to infinite-range behavior for a one-dimensional 
spin glass with long-range power law interactions \cite{katzgraber:04c}. For 
uncorrelated random variables, when the power of the exponent in the 
aforementioned model is large, the central limit theorem holds and a 
Gaussian distribution can be expected. But when the interactions are 
infinite-ranged and the variables are strongly correlated, such as in the 
case of the SK model, deviations from the central limit theorem can be 
expected, and, as previously mentioned, the data seem to follow a 
modified Gumbel distribution with a slope parameter 
expected to be \cite{palassini:03a} $m \approx 6$ [see Eq.~(\ref{modgumbel})]. 
This result is rather puzzling: What is the relationship between the 
ground-state energy distribution of the SK model and extreme value statistics? 
Recently Bertin and Clusel \cite{bertin:06} analytically derived the 
relationship between extreme value statistics and random sums of correlated
variables. Thus it might be plausible that similar arguments could be applied
to the SK model in order to explain the skewed energy distributions. In
particular, the results of Bertin and Clusel show for sums of correlated 
variables that there is not necessarily an underlying extreme process in order
to obtain a modified extreme value distribution. This means that the 
parameter $m$, expected to be integer when studied from the context of 
extreme value statistics, can also be noninteger, as has been found by
by Bramwell {\em et al.}~\cite{bramwell:00,bramwell:01} when studying
large-scale critical fluctuations in correlated systems.

Traditionally \cite{palassini:03a,katzgraber:04c}, simple-sampling techniques,
where $N_{\rm samp}$ disorder realizations are computed and subsequently binned
in order to obtain the ground-state energy distribution, are used. If $N_{\rm
samp}$ samples are computed, then the maximal ``resolution'' of a bin is $\sim
1/N_{\rm samp}$, and thus $\sim 10^6$ samples have to be computed to resolve six
orders of magnitude in the histogram. Therefore, the probing of tails in
ground-state energy distributions becomes quickly intractable making it
difficult to determine if a given fitting function properly describes the
tails of a ground-state energy distribution, as is the case for the SK model.
Multicanonical methods \cite{berg:91,berg:92} have been used before to 
overcome the limitations of simple-sampling techniques in order to probe
tails of overlap distribution functions in spin 
glasses \cite{berg:02,berg:02a}.
In this paper we outline a simple algorithm related to multicanonical
approaches based on ideas presented in
Ref.~\onlinecite{hartmann:02e} that also overcomes the limitations of
simple-sampling techniques by performing 
an importance-sampling simulation of the ground-state energy distribution 
\textit{in the disorder} with a {\em guiding function} 
computed from simple-sampling 
simulations. Similar approaches have been used before in the studies of
distributions of sequence alignment scores~\cite{hartmann:02e}, free-energy
barriers in the Sherrington-Kirkpatrick model \cite{bittner:06}, as well as 
fluctuations in classical magnets~\cite{hilfer:03} (albeit the latter
without disorder).

By computing the tails of the disorder distribution of the SK model to up to
18 orders of magnitude we show that a modified Gumbel distribution fits the
data well, yet with small systematic deviations. If the ground-state energy
distribution in the thermodynamic limit is a modified Gumbel distribution, 
then the slope parameter of the Gumbel distribution is considerably larger 
($m \approx 11$) than found in previous studies using simple-sampling 
techniques \cite{palassini:03a}.

We outline the method in Section~\ref{sec:method} and apply it to the
ground-state energy distribution of the SK model in
Section~\ref{sec:application}. We conclude in Sec.~\ref{sec:conclusion} and 
discuss the implications of our results on studies of ground-state 
energy distributions.

\section{Simulation of ground-state energy distributions}
\label{sec:method}

A disordered system is defined by a Hamiltonian
$\mathcal{H}_\mathcal{J}(\mathcal{C})$, where the disorder configuration
$\mathcal{J}$ is chosen from a probability distribution $P(\mathcal{J})$ and
$\mathcal{C}$ denotes the phase-space configuration of the system. The
ground-state energy $E$ of a given disorder 
configuration $\mathcal{J}$ is defined by
\begin{equation}
E(\mathcal{J}) = \min_{\mathcal{C}} \, \mathcal{H}_\mathcal{J}(\mathcal{C}).
\label{eminop}
\end{equation}
Together with the disorder distribution ${\mathcal P}(\mathcal{J})$, 
this defines the ground-state energy distribution
\begin{equation}
P(E) = \int d{\mathcal J} \, {\mathcal P}(\mathcal{J}) \, 
\delta \left[ E - E(\mathcal{J}) \right].
\end{equation}
To study the thermodynamic limit of the ground-state energy distribution, 
we use the standardized form $P_\mathrm{s}(x)$ defined by the equation
\begin{equation}
P(E) = \frac{1}{\sigma_E} P_\mathrm{s} 
\left( \frac{E-[E]_\mathrm{av}}{\sigma_E} \right),
\label{standard}
\end{equation}
where $[E]_\mathrm{av}$ and $\sigma_E = ( [E^2]_\mathrm{av} - [E]^2_\mathrm{av}
)^{1/2}$ are the average and the standard deviation of the
ground-state energy, respectively \cite{press:95}. Here $[\cdots]_{\rm av}$ represents an  
average over the disorder.

\subsection{Simple sampling}
\label{sec:simple}

Because $P(E)$ cannot be determined analytically for most disordered systems,
approximative methods such as Monte Carlo simulations have to be used. A
standard approach is to use a simple-sampling Monte Carlo algorithm to study
$P(E)$. $N_\mathrm{samp}$ independent disorder configurations $\mathcal{J}_i$
are chosen from $P(\mathcal{J})$ and the ground-state energy is calculated for
each disorder configuration. The calculation of the ground-state energy in
itself is a difficult optimization problem and can often only be solved
approximatively \cite{hartmann:01,hartmann:04}. From the ground-state energies
of these disorder configurations, the ground-state energy distribution can be
estimated as
\begin{equation}
P(E) = \frac{1}{N_\mathrm{samp}} \sum_{i=1}^{N_\mathrm{samp}} \delta 
\left[ E - E(\mathcal{J}_i) \right],
\label{pexp}
\end{equation}
so that the averages of functions with respect to the disorder are replaced by
averages with respect to the $N_\mathrm{samp}$ random samples. The functional
form of the ground-state energy distribution and its parameters can for example
be estimated by a maximum likelihood fit of a distribution $F_\theta(E)$ with
parameters $\theta$ to the data~\cite{cowan:98}. It corresponds to the
distribution for which the observed data have the largest probability. Due to
the limited range of energies sampled by the simple-sampling algorithm it is
often difficult or even impossible to quantify how well the tails of the 
distribution are described by a maximum-likelihood fit, and improving the data
by increasing the number of random samples is generally computationally 
very expensive. Therefore other methods have to be used to study the tails of
the distribution, and we outline in the next section a simple method that 
can be applied if a good analytic fit to the data can be found.

\subsection{Importance sampling with a guiding function}
\label{sec:is}

Assuming that we find a function $F_{\theta}(E)$ which accurately
describes the ground-state energy distribution as calculated with a
simple-sampling simulation, we now proceed by sampling the ground-state
distribution with an importance-sampling Monte Carlo algorithm \textit{in the
disorder}~\cite{newman:99,landau:00,hartmann:02e} and use
$F_{\theta}(E)$ as a guiding function \cite{pt_alex}. We start from a random
disorder configuration $\mathcal{J} = \mathcal{J}_0$ with ground-state energy
$E(\mathcal{J}_0)$. From the $i$-th configuration $\mathcal{J}_i$, we generate
the $i+1$-th configuration $\mathcal{J}_{i+1}$ by the following 
Metropolis-type \cite{metropolis:49} update:
\begin{enumerate}

\item Choose a candidate disorder configuration $\mathcal{J}'$ by replacing 
a subset of $\mathcal{J}$ chosen at random (e.g., a single bond chosen 
at random) with values chosen according to $P(\mathcal{J})$ [this requires
that $P(\mathcal{J})$ can be written in a product form]
and calculate its ground-state energy $E(\mathcal{J}')$.

\item Set $\mathcal{J}_{i+1} = \mathcal{J}'$ with probability
\begin{equation}
P_\mathrm{accept} = \min \left\{ 
\frac{F_{\theta}
\left[E(\mathcal{J}_i)\right]}{F_{\theta}\left[E(\mathcal{J'})\right]}
, 1 \right\}
\label{paccept}
\end{equation}
and $\mathcal{J}_{i+1} = \mathcal{J}_i$ otherwise.

\end{enumerate}
Using the importance-sampling Monte Carlo algorithm, a
disorder configuration $\mathcal{J}$ is visited with probability
$1/F_{\theta}[E(\mathcal{J})]$, such that the probability to visit a disorder
configuration with ground-state energy $E$ is $P(E)/F_{\theta}(E)$. If
$F_{\theta}(E) = P(E)$, then each energy is visited with the same probability
resulting in a flat-histogram sampling of the ground-state energy distribution.
To be able to study a finite range of energies and to avoid the trapping of
the algorithm in an extremal region of the energy space, the range of energies that the
algorithm is allowed to visit can be restricted. 

The main difference to the simple-sampling algorithm described in Section~\ref{sec:simple}
is that successive configurations visited by the algorithm are not independent.
Therefore an analysis of the results requires a quantification of the
correlations among configurations visited by the algorithm. This is a standard
problem in Markov Chain Monte Carlo simulations, and can for example be solved
with the help of the exponential autocorrelation time $\tau$ of the energy, 
which is the number of Monte Carlo steps after which the autocorrelation function 
of the ground-state energy
\begin{equation}
\chi(\Delta i) = \frac{\langle E_i^{} \, E_{i+\Delta i}^{} \rangle - 
\langle E_i^{} \rangle \, \langle E_{i+\Delta i}^{} \rangle}{\langle E_i^2 \rangle - 
\langle E_i^{} \rangle_{}^2},
\label{actime}
\end{equation}
decays to $1/e$ \cite{landau:00}. Here $E_i$ is the ground state energy after
the $i$-th Monte Carlo step and $\langle \ldots \rangle$ refers to the average
over Monte Carlo time. In order to ensure that the visited ground-state
configurations are not correlated, we only use every $4\tau$-th
measurement. Once the autocorrelation effects have been quantified, the data
can be analyzed with the same methods as the simple-sampling results 
[see Eq.~(\ref{pexp})].

\section{Application: Sherrington-Kirkpatrick model}
\label{sec:application}

The Sherrington-Kirkpatrick \cite{sherrington:75} model is defined by the
Hamiltonian
\begin{equation}
\mathcal{H}_\mathcal{J}(\{S_i\}) = \sum_{ i < j } J_{ij} \, S_i \, S_j,
\label{skham}
\end{equation}
where the $S_i = \pm 1$ ($i=1,\ldots,N$) are Ising spins, and the 
bonds $\mathcal{J} = \{J_{ij}\}$ are identically and independently 
distributed random variables chosen from a Normal distribution
with zero mean and standard deviation $(N-1)^{-1/2}$. The sum is over all
spins in the system.

The ground-state energy
of a given disorder configuration $\mathcal{J}$ is defined as
\begin{equation}
E(\mathcal{J}) = \min_{\{S_i\}} \, \mathcal{H}_\mathcal{J}(\{S_{i}\}).
\end{equation}
We are interested in the thermodynamic limit of the ground-state energy
distribution and define the limiting distribution $P_\infty(x)$ in analogy to
Eq.~(\ref{standard}) by
\begin{equation}
P(E) = \frac{1}{\sigma_E} P_\infty 
\left( \frac{E-[E]_\mathrm{av}}{\sigma_E} \right).
\end{equation}
For the SK model several optimization algorithms, such es extremal optimization
\cite{boettcher:01}, hysteretic optimization \cite{pal:06}, as well as other
algorithms such as genetic and Bayesian algorithms
\cite{hartmann:01,hartmann:04}, and even parallel tempering
\cite{marinari:92,hukushima:96,moreno:03,katzgraber:04c} are available. 
In this work we compute the
ground states of the system for the sake of simplicity using parallel tempering
Monte Carlo \cite{pt}. 
Note that a combination of the proposed importance-sampling
simulation in the disorder with other more efficient algorithms might yield 
more detailed results.
%%%%%%%
% table
\begin{table}
\caption{Maximum-likelihood fit to the simple-sampling Monte Carlo data 
calculated for the SK model in Ref.~\onlinecite{katzgraber:04c}. For each 
system size $N$, $N_\mathrm{samp} = 10^5$ random samples have been generated. 
The maximum-likelihood fit has been performed by binning the data into 
50 bins. The error bars and the parametric estimates of $[E]_\mathrm{av}$ 
and $\sigma_E$ [see Eqs.~(\ref{rel1}) and (\ref{rel2})] have been generated 
by a bootstrap method (see text).}
\label{simsamptable}
\begin{tabular*}{\columnwidth}{@{\extracolsep{\fill}} c l l l l l}
\hline
\hline
$N$  &  $\mu$  & $\nu$ & $m$ & 
$[E]_\mathrm{av}$\footnotemark[1] \footnotetext[1]{Parametric estimate 
calculated from $\mu$, $\nu$, and $m$.}
&
$\sigma_E$\footnotemark[1]  \\
\hline
$16$ & $-10.373(6)$ & $2.48(4)$ & $4.92(16)$ & $-10.634(4)$ & $1.180(3)$ \\
$24$ & $-16.189(6)$ & $2.81(5)$ & $4.99(17)$ & $-16.481(4)$ & $1.326(3)$ \\
$32$ & $-22.076(6)$ & $3.23(6)$ & $5.59(19)$ & $-22.375(5)$ & $1.432(3)$ \\
$48$ & $-33.935(8)$ & $3.81(8)$ & $6.24(26)$ & $-34.249(5)$ & $1.590(4)$ \\
$64$ & $-45.848(10)$ & $3.97(9)$ & $5.87(26)$ & $-46.196(5)$ & $1.712(4)$ \\
$96$ & $-69.840(9)$ & $4.53(9)$ & $6.37(25)$ & $-70.205(6)$ & $1.867(5)$ \\
$128$ & $-93.918(13)$ & $4.89(14)$ & $6.49(39)$ & $-94.305(6)$ & $1.997(5)$ \\
\hline
\hline
\end{tabular*}
\end{table}
%
%%%%%%%
We start by studying the $10^5$ ground-state energies calculated for
different system sizes in Ref.~\onlinecite{katzgraber:04c}. We bin the data into 50
bins and perform a maximum-likelihood fit with a modified Gumbel
distribution \cite{gumbel:60,palassini:03a} which seems to fit the
simple-sampling data well
\begin{equation}
G_{\mu,\nu,m}(E) \propto \exp \left[ m \,
\frac{E-\mu}{\nu}  - m \, \exp \left( \frac{E-\mu}{\nu} \right)
\right].
\label{modgumbel}
\end{equation}
The modified Gumbel distribution is parameterized by the ``location'' parameter
$\mu$, the ``width'' parameter $\nu$, and the ``slope'' parameter $m$. To
obtain error bars we generate 200 bootstrap replicates~\cite{efron:94} of the data
and perform the maximum-likelihood fit for each bootstrap replicate. Because the
average and standard deviation of the modified Gumbel distribution are related
to the parameters $\mu$, $\nu$, and $m$ by
\begin{eqnarray}
[E]_\mathrm{av} & = & \mu + E_m \, \nu \label{rel1} \\
\sigma_E & = & \sigma_m \, \nu, \label{rel2}
\end{eqnarray}
where $E_m$ and $\sigma_m$ are the average and standard deviation of
$G_{0,1,m}$, a parametric estimate of the average energy and its standard
deviation can be calculated for the data (and each bootstrap replicate). The
results of the analysis are summarized in Table~\ref{simsamptable}.
Note that for the simple-sampling data we find $m \approx 6$ as suggested in
Ref.~\onlinecite{palassini:03a} and that the parametric estimates of
$[E]_\mathrm{av}$ and $\sigma_E$ agree within error bars with the direct
estimates calculated in Ref.~\onlinecite{katzgraber:04c} using Eqs.~(4) and (5) 
therein.

\subsection*{Importance-sampling simulation}
\label{sec:applicationmulti}

%%%%
% parameter table
\begin{table}
\caption{Input parameters for the guiding function used in the simulation.
Initial runs using the parameters shown in Table~\ref{simsamptable} indicated
that $m > 6$, so that we have chosen $m = 8$ for the production runs. The
parameters $\mu$ and $\nu$ have been calculated from $[E]_\mathrm{av}$, $\sigma_e$,
and $m$ with the help of Eqs.~(\ref{rel1}) and (\ref{rel2}). \label{partable}}
\begin{tabular*}{\columnwidth}{@{\extracolsep{\fill}} c l l l l l}
\hline
\hline
$N$  &  $[E]_\mathrm{av}$ & $\sigma_e$ & $m$ & $\mu$  & $\nu$ \\
\hline
$16$ & $-10.635$ & $1.180$ & $8$ & $-10.429$ & $3.233$ \\
$24$ & $-16.481$ & $1.325$ & $8$ & $-16.249$ & $3.631$ \\
$32$ & $-22.321$ & $1.469$ & $8$ & $-22.064$ & $4.025$ \\
$48$ & $-34.248$ & $1.589$ & $8$ & $-33.970$ & $4.355$ \\
$64$ & $-70.205$ & $1.710$ & $8$ & $-45.961$ & $4.685$ \\
$96$ & $-94.305$ & $1.867$ & $8$ & $-69.789$ & $5.116$ \\
$128$ & $-142.627$ & $2.186$ & $8$ & $-93.956$ & $5.472$ \\
\hline
\hline
\end{tabular*}
\end{table}
\begin{figure}
\begin{center}
\includegraphics[width=7.8cm]{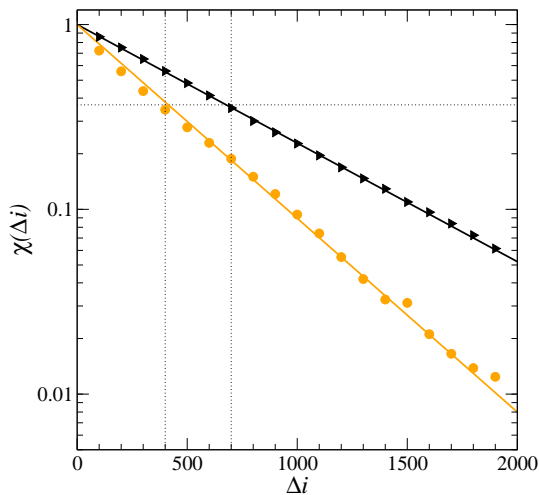}
\end{center}
\caption{Autocorrelation function as defined in Eq.~(\ref{actime}) for system
sizes $N=16$ (circles) and $N=128$ (triangles) for the simulation with 
parameters shown in Table~\ref{partable}. The value $1/e$ is marked by 
the horizontal dotted line ($\Delta i$ is measured in Monte Carlo steps).}
\label{actimefig}
\end{figure}
\begin{figure}
\begin{center}
\includegraphics[width=7.0cm]{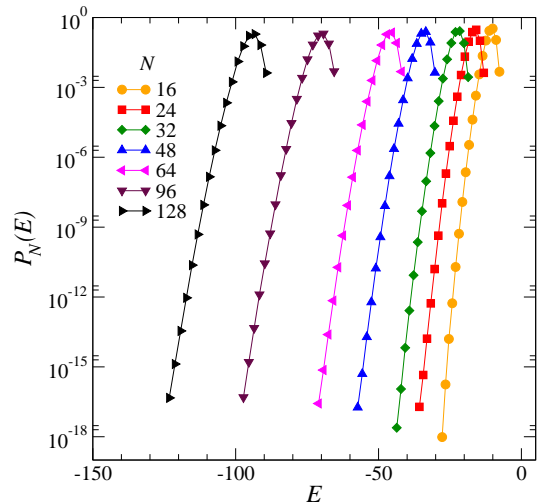}
\end{center}
\caption{Unscaled ground-state energy distributions of the
Sherrington-Kirkpatrick model for different system sizes, obtained by the
guiding-function simulation with the parameters given in Table~\ref{partable}.
For each system size, between $92$ and $686$ independent samples are simulated.}
\label{unscaleddata}
\end{figure}
\begin{figure}
\begin{center}
\includegraphics[width=7.0cm]{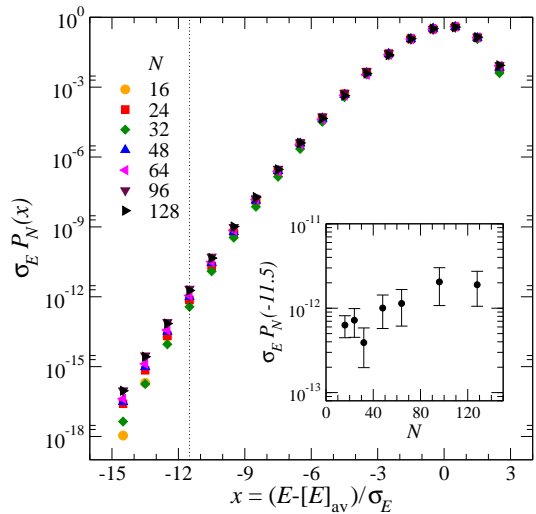}
\end{center}
\caption{Scaled ground-state energy distributions of the
Sherrington-Kirkpatrick model for different system sizes. The curves are 
scaled by the average energy and standard deviation obtained from
simple-sampling simulations shown in Table~\ref{simsamptable}. The scaled
distributions obtained for different system sizes collapse on a common curve,
although finite size effects are clearly visible in the tail. The inset shows
the behavior of the ground-state energy distribution at $x = -11.5$ (vertical 
dotted line in the main panel) as a function of the system size $N$ in order
to illustrate finite-size effects.
}
\label{scaleddata}
\end{figure}
\begin{table}
\caption{Three-parameter fit in the parameters $\mu$,
$\nu$, and $m$ to the data [rescaled to $x = (E-[E]_{\rm av})/\sigma_{E}$]
for the SK model. For each system size $N$, between $N_\mathrm{samp} = 92$ 
and $686$ independent samples have been generated. $z=\nu/m$ describes the 
asymptotic behavior of the single-exponential tail (error bar obtained from
independent fits). $\chi^2/{\rm dof}$ represents the 
$\chi^2$ per degree of freedom of the fit \cite{press:95}.
\label{threeparfit}}
\begin{tabular*}{\columnwidth}{@{\extracolsep{\fill}} c r r l l l c}
\hline
\hline
$N$  & $N_\mathrm{samp}$ & $\quad\mu\hfill$  & $\nu$ & $m$ & $\nu/{m}$ & $\chi^2/{\rm dof}$\\
\hline
$16$ & $686$ & $-0.059(63)$ & $3.87(32)$ & $13.8(16)$ & $0.279(9)$ & $6.29$\\
$24$ & $274$ & $-0.024(41)$ & $3.55(19)$ & $11.9(9)$ & $0.298(6)$  & $1.04$\\
$32$ & $311$ & $-0.017(43)$ & $3.31(18)$ & $11.2(9)$ & $0.295(6)$  & $1.47$\\
$48$ & $221$ & $ 0.010(40)$ & $3.50(17)$ & $11.5(8)$ & $0.303(5)$  & $0.77$\\
$64$ & $168$ & $ 0.063(44)$ & $3.42(18)$ & $11.1(8)$ & $0.309(5)$  & $0.72$\\
$96$ & $92$ &  $ 0.026(36)$ & $3.44(14)$ & $10.9(6)$ & $0.314(4)$  & $0.23$\\
$128$ & $112$ & $0.066(42)$ & $3.39(16)$ & $10.7(7)$ & $0.317(5)$  & $0.42$\\
\hline
\hline
\end{tabular*}
\end{table}
\begin{figure}
\begin{center}
\includegraphics[width=7.0cm]{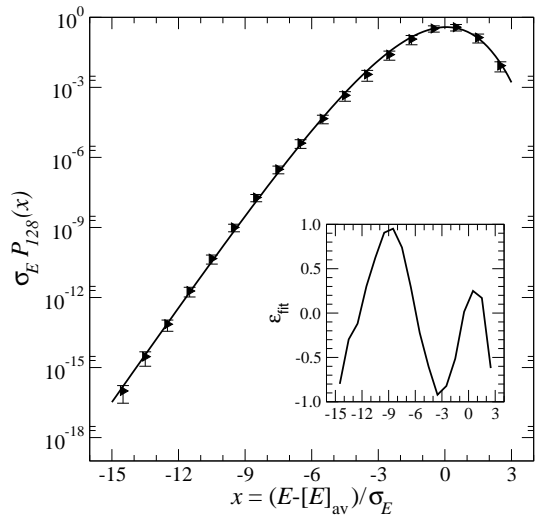}
\end{center}
\caption{Scaled ground-state energy distribution for $N = 128$ and the
fit to the modified Gumbel distribution as described in the text. 
The inset shows 
$\epsilon_\mathrm{fit} = (P_\mathrm{fit} - P_{128})/\Delta P_{128}$, the fit
deviation normalized by the error of the calculated ground-state energy
distribution.} 
\label{fitdata}
\end{figure}
Using the maximum-likelihood estimates shown in Table~\ref{simsamptable}, we
now perform an importance-sampling simulation in the disorder as described in
Section~\ref{sec:is}. To perform a step in the Monte
Carlo algorithm, we choose a site at random, replace all bonds connected to 
this site (the expected change in the ground-state energy is then of the order $\sim 1/N$), 
calculate the ground-state energy of the new configuration, and
accept the new configuration with the probability given in 
Eq.~(\ref{paccept}), which in this case is a modified Gumbel distribution,
Eq.~(\ref{modgumbel}), with the parameters listed in Table \ref{partable}.
To avoid a trapping of the simulation in the double-exponential
forward tail of the distribution, we limit the maximum energy allowed in the
simulation by $[E]_\mathrm{av} + 3 \sigma_E$. Initial runs indicated that 
the estimates of $m$ shown in Table~\ref{simsamptable} are too small, and we
therefore choose an estimate of $m = 8$ for the production runs. $\mu$ and $\nu$ are determined 
from the simple-sampling results for the average and standard
deviation with the help of Eqs.~(\ref{rel1}) and (\ref{rel2}), and a summary
of the input parameters used is shown in Table~\ref{partable}.
It is important to note that this change in the parameters does not lead
to a systematic error or bias in the results and merely constitutes a change of 
the guiding function, which can either improve or degrade the range of
energies visited by the algorithm.
Figure~\ref{actimefig} shows the energy-energy autocorrelation function for system 
sizes $N=16$ and $128$. Autocorrelation times are of the order of $400$ to $700$ 
Monte Carlo steps resulting in $92$ to $686$ independent measurements for the different
system sizes. 
While the number of samples used is small, the method
is able to probe the tails in this particular case down to $18$ orders of
magnitude, a result {\em impossible} to obtain with simple-sampling
techniques.
Figure~\ref{unscaleddata} shows the unscaled ground-state energy distributions
obtained from the simulation. Figure~\ref{scaleddata} shows the
same distributions standardized with the average energy and the standard deviation
as measured from simple-sampling simulations. We choose to standardize with 
$[E]_\mathrm{av}$ and $\sigma_E$ from the simple-sampling simulations because of the 
greater number of statistically -- for these quantities -- relevant measurements. 
The standardized distributions obtained for different system sizes collapse on a single
curve, although finite size effects are clearly visible in the tail of the distribution (see 
also inset of Figure~\ref{scaleddata}).

Next we try to determine the functional form of the distribution by fitting
a modified Gumbel distribution to the data. We have used the standard
fitting function of the {\tt gnuplot} package, which implements
the nonlinear least-squares (NLLS) Marquardt-Levenberg algorithm.
The results of the fit are shown as $\mu$, $\nu$, and $m$ in 
Table~\ref{threeparfit}. In particular, the values for $m$ are 
approximately two times larger than previous results \cite{palassini:03a}. 

Figure~\ref{fitdata} shows the ground-state energy distribution for $N=128$
together with the resulting fit. The data have been standardized
by the simple-sampling results and fitted with a modified Gumbel
distribution.
To examine the systematic deviations more closely, the inset shows the 
deviation $\epsilon_{\rm fit}$
of the fit from the observed data scaled by the measurement
error $\Delta P$ obtained from a bootstrap analysis. If no systematic
deviations were present, one would expect the deviations to be of order
one with an irregularly changing sign. Indeed $\epsilon_{\rm fit}$ varies 
between $-1$ and $1$, showing the good quality of the fit. Nevertheless, 
the differences between fit and data have clear systematic deviations, which 
might suggest that a modified Gumbel distribution is not the correct 
asymptotic probability distribution function. Note that we also
excluded by a visual comparison that the distribution has the form of a 
Tracy-Widom distribution~\cite{andreanov:04,rmlab}.

\begin{figure}
\begin{center}
\includegraphics[width=7.0cm]{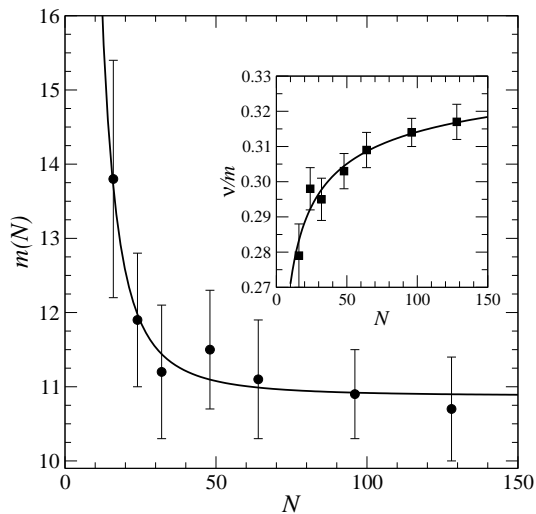}
\end{center}
\caption{Finite-size dependence of the parameter $m$ 
[see Eq.~(\ref{modgumbel})] from the fits to the modified Gumbel 
distribution, together with a power-law fit (see text). The inset shows 
a similar analysis for the tail parameter $z=\nu/m$. The data converge to 
$m_\infty = 10.9(5)$ and $z_\infty = 0.34(4)$, respectively.
} 
\label{fig:mzN}
\end{figure}

Because the systematic deviations between fitted function and data decrease 
with system size (not shown) and the fitted function agrees relatively well
with the data over several orders of magnitude, it can be surmised 
empirically that a modified Gumbel distribution might present a good
description of the limiting distribution function for the SK model.
Thus we attempt to determine the asymptotic behavior of the tail by 
extrapolating to infinite system size by fitting a power law
$m(N)=m_{\infty}+aN^{-b}$ to the data for the values of $m$ shown in Table 
\ref{threeparfit}. The resulting limiting value is 
\begin{equation}
m_{\infty}=10.9(5),
\end{equation}
where for the fit $\chi^2/{\rm dof} = 0.331$.
The data points together with the resulting fit are shown in 
Fig.~\ref{fig:mzN}. The inset shows a similar analysis for $z=\nu/m$ which 
describes the behavior in the exponential tail of the Gumbel distribution, 
resulting in $z_\infty=0.34(4)$. 

\section{Conclusions}
\label{sec:conclusion}
In this paper we have explained an importance-sampling algorithm with
a guiding function to simulate 
the ground-state energy distribution of a disordered system to high order
using a {\em considerably smaller} numerical effort than with simple-sampling 
techniques. When compared to full multicanonical simulation schemes to sample
distributions, such as used in Ref.~\onlinecite{hartmann:02e} and 
Ref.~\onlinecite{hilfer:03}, our algorithm has several advantages due to 
its simplicity: Instead of iterating towards a good guiding function, which 
may be quite expensive computationally, we use a maximum likelihood fit as a 
guiding function. Therefore, the algorithm we propose is straightforward to 
implement and considerably more efficient than traditional approaches,
provided a good guiding function, i.e., a good maximum-likelihood fit to the
simple-sampling results, can be found. Note also that the method can be
generalized to any distribution function, such as an order-parameter
distribution.

We have illustrated the algorithm with the computation of the ground-state 
energy distribution of the Sherrington-Kirkpatrick model and find that
the ground-state energy distribution can be described over several orders of
magnitude by a modified Gumbel distribution albeit with systematic deviations.
Therefore, if the limiting probability distribution ($N \rightarrow \infty$)
of the SK model is a modified Gumbel distribution, it has a slope 
parameter $m\approx 11$, a value significantly larger than estimated before. 
Simulations with more efficient ground-state search algorithms in order to
probe larger system sizes would be desirable in order to see if the
aforementioned systematic deviations become negligible for large $N$.
Note that our results for the mean energy
$E_{\rm av}$ as well as the fluctuations $\sigma_{E}$ are not influenced by
the importance sampling technique because the method probes the tails where
probabilities are small and thus contributions to the moments are negligible. 

Since the method can be applied more generally to systems where
simple-sampling results exist, revisiting the one-dimensional Ising 
chain with random power-law interactions \cite{katzgraber:04c} together with
a better ground-state search algorithm would be desirable in order to probe
the crossover from mean-field to non-mean-field behavior in more detail.

\begin{acknowledgments}
We would like to thank A.~P.~Young for comments on the manuscript.
We also thank M.~Troyer for his input and C.~Tracy and M.~Dieng for 
providing the Mathematica and Matlab code~\cite{rmlab} to calculate the 
Tracy-Widom distribution. The simulations have been performed on the 
Hreidar cluster at ETH Z\"urich. A.~K.~Hartmann has been funded by the 
{\em VolkswagenStiftung} (Germany) within the program ``Nachwuchsgruppen 
an Universit\"aten'' and by the DYGLAGEMEM contract
(HPRN-CT-2002-00307) of the European community.
\end{acknowledgments}

\bibliography{refs,comments}

\begin{thebibliography}{39}
\bibitem[{pt_()}]{pt_alex}
\bibinfo{note}{Note that the method outlined here has advantages over the
  method using parallel tempering presented in Ref.~\onlinecite{hartmann:02e}:
  Parallel tempering fails when the logarithm of the energy distribution,
  $\log[P(E)]$, is not concave. Since $d \log[P(E)]/dE$ corresponds to an
  inverse effective temperature at which the energy $E$ is sampled, different
  points in the distribution would correspond to the same temperature if the
  logarithm of the distribution is not concave. Thus parallel tempering would
  be difficult to tune. A prime example of a system where $\log[P(E)]$ is not
  concave is when a first-order phase transition is present. In that case
  $P(E)$ has a two-peak structure.}
\bibitem[{pt()}]{pt}
\bibinfo{note}{Details of the parallel-tempering Monte Carlo ground-state
  search as well as simulation parameters used to compute the ground-state
  energies of the Sherrington-Kirkpatrick model are presented in Ref.
  \onlinecite{katzgraber:04c}, in particular see Appendix A.}
\end{thebibliography}

\end{document}